\documentclass{ws-procs975x65}

\begin{document}

\title{RECENT ADVANCES IN THE THEORY OF NUCLEAR FORCES
AND ITS IMPACT ON MICROSCOPIC NUCLEAR
STRUCTURE\footnote{Talk presented at EXOCT 2007, Catania, 
Italy, June 11-15, 2007.}}

\author{R. Machleidt}

\address{Department of Physics, University of Idaho,\\
Moscow, Idaho 83844, U.S.A.\\
E-mail: machleid@uidaho.edu}

\begin{abstract}
The theory of nuclear forces has made great progress since the turn
of the millenium using the framework of chiral effective field theory
(ChEFT).
The advantage of this approach, which was originally proposed
by Weinberg, is that it has a firm basis in
quantum-chromodynamics and allows for quantitative 
calculations.
Moreover, this theory generates 
two-nucleon forces (2NF) 
and many-body forces on an equal footing and provides an explanation
for the empirically known fact that 2NF $\gg$ 3NF $\gg$ 4NF.
I will present the recent advances in more detail and put them into
historical context.
In addition, I will also provide a critical evaluation
of the progress made including a discussion of the limitations 
of the ChEFT approach.
\end{abstract}


\bodymatter

\section{Introduction and historical perspective}

The theory of nuclear forces has a long history (cf.\ Table~1).
Based upon the seminal idea by Yukawa~\cite{Yuk35}, 
first field-theoretic attempts
to derive the nucleon-nucleon (NN) interaction
focused on pion-exchange.
While the one-pion exchange turned out to be very useful
in explaining NN scattering data and the properties
of the deuteron, 
multi-pion exchange was beset with
serious ambiguities.
Thus, the ``pion theories'' of the 1950s
are generally judged as failures---for reasons
we understand today: pion dynamics is constrained by chiral
symmetry, a crucial point that was unknown in the 1950s.

\begin{table}[t]
\tbl{Seven Decades of Struggle:
The Theory of Nuclear Forces} 
{\begin{tabular*}{\textwidth}{@{\extracolsep{\fill}}cccc}
\hline
\hline
\\
   & 
  \bf 1935   &
\bf Yukawa: Meson Theory &
\\
\\
\hline
     &      &
{\it The ``Pion Theories''}
\\
 & \bf 1950's &
One-Pion Exchange: o.k.
\\
  &         &
Multi-Pion Exchange: disaster
\\
\hline
  &         & 
Many pions $\equiv$ multi-pion resonances:
\\
 & \bf 1960's & 
{\boldmath $\sigma$, $\rho$, $\omega$, ...}
\\
  &         & 
The One-Boson-Exchange Model
\\
\hline
  &         &
            Refine meson theory:
\\
 & \bf 1970's & 
Sophisticated {\boldmath $2\pi$} exchange models
\\
  &         & 
(Stony Brook, Paris, Bonn)
\\
\hline
  &         & Nuclear physicists discover
\\
 & \bf 1980's &  {\bf QCD}
\\
  &        & Quark Cluster Models
\\
\hline
  &        &
Nuclear physicists discover {\bf EFT}
\\
 & \bf 1990's &
Weinberg, van Kolck
\\
 & \bf and beyond &
{\bf Back to Meson Theory!}
\\
  &           &
{\it But, with Chiral Symmetry}
\\
\hline
\hline
\end{tabular*}}
\end{table}

Historically, the experimental discovery of heavy 
mesons in the early 1960s
saved the situation. The one-boson-exchange (OBE)
model~\cite{Mac89} emerged which is still the most economical
and quantitative
phenomenology for describing the 
nuclear force~\cite{Sto94,Mac01}.
The weak point of this model, however, is the scalar-isoscalar
``sigma'' or ``epsilon'' boson, for which the empirical
evidence remains controversial. Since this boson is associated
with the  correlated (or resonant) exchange of two pions,
a vast theoretical effort that occupied more than a decade 
was launched to derive the 2$\pi$-exchange contribution
to the nuclear force, which creates the intermediate 
range attraction.
For this, dispersion theory as well as 
field theory were invoked producing  the
Paris~\cite{Lac80} and the Bonn~\cite{Mac89,MHE87}
potentials.

The nuclear force problem appeared to be solved; however,
with the discovery of quantum chromo-dynamics (QCD), 
all ``meson theories'' were
relegated to models and the attempts to derive
the nuclear force started all over again.

The problem with a derivation from QCD is that
this theory is non-perturbative in the low-energy regime
characteristic of nuclear physics, which makes direct solutions
impossible.
Therefore, during the first round of new attempts,
QCD-inspired quark models~\cite{MW88} became popular. 
These models are able to reproduce
qualitatively and, in some cases, semi-quantitatively
the gross features of the nuclear 
force~\cite{EFV00,Wu00}.
However, on a critical note, it has been pointed out
that these quark-based
approaches are nothing but
another set of models and, thus, do not represent any
fundamental progress. Equally well, one may then stay
with the simpler and much more quantitative meson models.

A major breakthrough occurred when 
the concept of an effective field theory (EFT) was introduced
and applied to low-energy QCD~\cite{Wei79}.

Note that the QCD Lagrangian for massless up and down quarks
is chirally symmetric, i.~e., it is invariant under
global flavor 
$SU(2)_L \times SU(2)_R$ equivalent to
$SU(2)_V \times SU(2)_A$ (vector and axial vector)
transformations. The axial symmetry is spontaneously broken
as evidenced in the absence of parity doublets in the
low-mass hadron spectrum. This implies the existence 
of three massless Goldstone bosons which are identified
with the three pions ($\pi^\pm, \pi^0$).
The non-zero, but small, pion mass is a consequence of
the fact that the up and down quark masses are not
exactly zero either (some small, but explicit symmetry breaking).
Thus, we arrive at a low-energy scenario 
that consists 
of pions and nucleons interacting via a force
governed by spontaneously broken approximate chiral
symmetry. 

To create an effective field theory describing this scenario,
one has to write down the most general Lagrangian consistent
with the assumed symmetry principles, particularly
the (broken) chiral symmetry of QCD~\cite{Wei79}.
At low energy, the effective degrees of freedom are pions and
nucleons rather than quarks and gluons; heavy mesons and
nucleon resonances are ``integrated out''.
So, the circle of history is closing and we are back to Yukawa's meson theory,
except that we have learned to add one important refinement to the theory:
broken chiral symmetry is a crucial constraint that generates
and controls the dynamics and establishes a clear connection
with the underlying theory, QCD.

It is the purpose of the remainder of this contribution to describe
the EFT approach to nuclear forces in more detail.

\section{Chiral perturbation theory and the hierarchy
of nuclear forces}

The chiral effective Lagrangian 
is given by an infinite series
of terms with increasing number of derivatives and/or nucleon
fields, with the dependence of each term on the pion field
prescribed by the rules of broken chiral symmetry.
Applying this Lagrangian to NN scattering generates an unlimited
number of Feynman diagrams.
However, Weinberg showed~\cite{Wei90} that a systematic expansion
exists in terms of $(Q/\Lambda_\chi)^\nu$,
where $Q$ denotes a momentum or pion mass, 
$\Lambda_\chi \approx 1$ GeV is the chiral symmetry breaking
scale, and $\nu \geq 0$ (cf.\ Fig.~1).
This has become known as chiral perturbation theory ($\chi$PT).
For a given order $\nu$, the number of terms is
finite and calculable; these terms are uniquely defined and
the prediction at each order is model-independent.
By going to higher orders, the amplitude can be calculated
to any desired accuracy.

\begin{figure}[t]
\vspace*{-0.7cm}
\hspace*{0.8cm}
\psfig{file=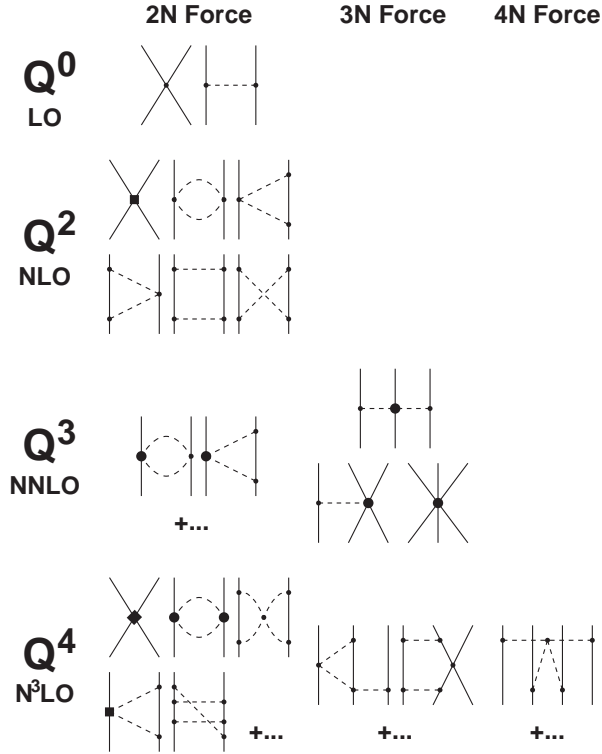,width=90mm}
\vspace*{-0.7cm}
\caption{Hierarchy of nuclear forces in $\chi$PT. 
Solid lines represent nucleons and dashed lines pions. 
Further explanations are given in the text.}
\end{figure}

Following the first initiative by Weinberg \cite{Wei90}, pioneering
work was performed by Ord\'o\~nez, Ray, and
van Kolck \cite{ORK94,Kol99} who 
constructed a NN potential in coordinate space
based upon $\chi$PT at
next-to-next-to-leading order (NNLO; $\nu=3$).
The results were encouraging and
many researchers became attracted to the new field.
Kaiser, Brockmann, and Weise~\cite{KBW97} presented the first model-independent
prediction for the NN amplitudes of peripheral
partial waves at NNLO.
Epelbaum {\it et al.}~\cite{EGM98} developed the first momentum-space
NN potential at NNLO, and Entem and Machleidt~\cite{EM03} presented the first
potential at N$^3$LO ($\nu = 4$).

In $\chi$PT, the NN amplitude is uniquely determined
by two classes of contributions: contact terms and pion-exchange
diagrams. There are two contacts of order $Q^0$ 
[${\cal O}(Q^0)$] represented by the four-nucleon graph
with a small-dot vertex shown in the first row of Fig.~1.
The corresponding graph in the second row, four nucleon legs
and a solid square, represent the 
seven contact terms of ${\cal O}(Q^2)$. 
Finally, at ${\cal O}(Q^4)$, we have 15 contact contributions
represented by a four-nucleon graph with a solid diamond.

Now, turning to the pion contributions:
At leading order [LO, ${\cal O}(Q^0)$, $\nu=0$], 
there is only the wellknown static one-pion exchange, second
diagram in the first row of Fig.~1.
Two-pion exchange (TPE) starts
at next-to-leading order (NLO, $\nu=2$) and all diagrams
of this leading-order two-pion exchange are shown.
Further TPE contributions occur in any higher order.
Of this sub-leading TPE, we show only
two representative diagrams at NNLO and three diagrams at N$^3$LO.
The TPE at N$^3$LO has been calculated first by
Kaiser~\cite{Kai01}. All $2\pi$ exchange diagrams/contributions up to
N$^3$LO are summarized in a pedagogical and systematic
fashion in Ref.~\refcite{EM02} where 
the model-independent results for NN scattering in peripheral
partial waves are also shown. 

Finally, there is also three-pion exchange, which
shows up for the first time 
at N$^3$LO (two loops; one representative $3\pi$ diagram
is included in Fig.~1). 
In Ref.~\refcite{Kai99}
it was demonstrated that the 3$\pi$ contribution at this order
is negligible. 

One important advantage of $\chi$PT is that it makes specific
predictions also for many-body forces. For a given order of $\chi$PT,
two-nucleon forces (2NF), three-nucleon forces (3NF), 
\ldots are generated on the same footing (cf.\ Fig.~1). 
At LO, there are no 3NF, and
at next-to-leading order (NLO),
all 3NF terms cancel~\cite{Wei90,Kol94}. 
However, at NNLO and higher orders, well-defined, 
nonvanishing 3NF occur~\cite{Kol94,Epe02b}.
Since 3NF show up for the first time at NNLO, they are weak.
Four-nucleon forces (4NF) occur first at N$^3$LO and, therefore,
they are even weaker.

\section{Chiral NN potentials}

\begin{table}[t]
\tbl{$\chi^2$/datum for the reproduction of the 1999 $np$ database
by families of $np$ potentials at NLO and NNLO constructed by the
Bochum/Juelich group~\cite{EGM04}.}
{\begin{tabular*}{\textwidth}{@{\extracolsep{\fill}}cccccc}
\hline 
\hline 
 &  \# of $np$ && \multicolumn{3}{c}{\it Bochum/Juelich}\\
 Bin (MeV) 
 & data 
 &
 & NLO
 & 
 & NNLO 
\\
\hline 
\hline 
0--100&1058&&4--5&&1.4--1.9\\ 
100--190&501&&77--121&&12--32\\ 
190--290&843&&140--220&&25--69\\ 
\hline 
0--290&2402&&67--105&&12--27
\\ 
\hline 
\hline 
\end{tabular*}}
\end{table}

The two-nucleon system is non-perturbative as evidenced by the
presence of shallow bound states and large scattering lengths.
Weinberg~\cite{Wei90} showed that the strong enhancement of the
scattering amplitude arises from purely nucleonic intermediate
states. He therefore suggested to use perturbation theory to
calculate the NN potential (Fig.~1) 
and to apply this potential
in a scattering equation (Lippmann-Schwinger or Schr\"odinger 
equation) to obtain the NN amplitude. We follow 
this philosophy.

Chiral perturbation theory is a low-momentum expansion.
It is valid only for momenta 
$Q \ll \Lambda_\chi \approx 1$ GeV.
Therefore, when a potential is constructed, all expressions (contacts and
irreducible pion exchanges) are multiplied with a regulator function,
\begin{equation}
\exp\left[ 
-\left(\frac{p}{\Lambda}\right)^{2n}
-\left(\frac{p'}{\Lambda}\right)^{2n}
\right] \; ,
\end{equation}
where $p$ and $p'$ denote, respectively, the magnitudes
of the initial and final nucleon momenta in the center-of-mass
system (CMS); 
and $\Lambda \ll \Lambda_\chi$. The exponent $2n$ is to be chosen
such that 
the regulator generates powers which are beyond
the order at which the calculation is conducted.

To what order in $\chi$PT do we have to go for
sufficient accuracy?
To discuss this issue on firm grounds, I show in Table~2
the $\chi^2$/datum for the fit of the world $np$ data
below 290 MeV for a family of $np$ potentials at 
NLO and NNLO. 
The NLO potentials produce the horrendous $\chi^2$/datum between 67 and 105,
and the NNLO are between 12 and 27.
The rate of improvement from one order to the other
is very impressive, but the quality of the reproduction
of the $np$ data at NLO and NNLO is obviously totally
insufficient for reliable predictions.

\begin{table}[t]
\tbl{$\chi^2$/datum for the reproduction of the 1999 {\boldmath\bf $np$ database}
by various $np$ potentials.
Numbers in parentheses denote cutoff parameters in units of MeV.}
{\begin{tabular*}{\textwidth}{@{\extracolsep{\fill}}cc|c|c|c}
\hline 
\hline 
 &  
 & {\it Idaho}
 & {\it Bochum/Juelich}
 & Argonne         
\\
 Bin (MeV)
 & \# of {\boldmath $np$}
 & N$^3$LO~\cite{EM03}
 & N$^3$LO~\cite{EGM05} 
 & $V_{18}$~\cite{WSS95}
\\
 &  data
 & (500--600)
 & (600/700--450/500)
 & 
\\
\hline 
\hline 
0--100&1058&1.0--1.1&1.0--1.1&0.95\\ 
100--190&501&1.1--1.2&1.3--1.8&1.10\\ 
190--290&843&1.2--1.4&2.8--20.0&1.11\\ 
\hline 
0--290&2402&1.1--1.3&1.7--7.9&1.04
\\ 
\hline 
\hline 
\end{tabular*}}
\end{table}

\begin{table}[b]
\tbl{$\chi^2$/datum for the reproduction of the 1999 {\boldmath\bf $pp$ database}
by various $pp$ potentials.
Numbers in parentheses denote cutoff parameters in units of MeV.}
{\begin{tabular*}{\textwidth}{@{\extracolsep{\fill}}cc|c|c|c}
\hline 
\hline 
              
 &  
 & {\it Idaho}
 & {\it Bochum/Juelich}
 & Argonne         
\\
 Bin (MeV)
 & \# of {\boldmath $pp$}
 & N$^3$LO~\cite{EM03}
 & N$^3$LO~\cite{EGM05} 
 & $V_{18}$~\cite{WSS95}
\\
 &  data
 & (500--600)
 & (600/700--450/500)
 & 
\\
\hline 
\hline 
0--100&795&1.0--1.7&1.0--3.8&1.0 \\ 
100--190&411&1.5--1.9&3.5--11.6&1.3 \\ 
190--290&851&1.9--2.7&4.3--44.4&1.8 \\ 
\hline 
0--290&2057&1.5--2.1&2.9--22.3&1.4 
\\ 
\hline 
\hline 
\end{tabular*}}
\end{table}

Based upon these facts, it has been pointed out in 2002 by
Entem and Machleidt~\cite{EM02a,EM02} that NNLO is insufficient and one has
to proceed to N$^3$LO. Consequently, the first N$^3$LO  potential was
created in 2003~\cite{EM03}, which showed that at this order
a $\chi^2$/datum comparable to the high-precision
Argonne $V_{18}$ potential can, indeed, be achieved, see Tables~3 and 4.
This ``Idaho'' N$^3$LO potential~\cite{EM03} produces
a $\chi^2$/datum = 1.1 
for the world $np$ data below 290 MeV
which compares well with the $\chi^2$/datum = 1.04
by the Argonne potential (Table~3).
In 2005, also the Bochum/Juelich group produced
several N$^3$LO NN potentials~\cite{EGM05}, the best of which
fits the $np$ data with
a $\chi^2$/datum = 1.7 and the worse with 
a $\chi^2$/datum = 7.9 (cf.\ Table~3).
While 7.9 is clearly unacceptable for any meaningful
application, a $\chi^2$/datum of 1.7 may be sufficient for
most purposes.

I turn now to the $pp$ data, Table~4.
Typically, $\chi^2$ for $pp$ data are larger than for $np$
because of the higher precision of $pp$ data.
Thus, the Argonne $V_{18}$ produces
a $\chi^2$/datum = 1.4 for the world $pp$ data
below 290 MeV and the best Idaho N$^3$LO $pp$ potential obtains
1.5. The fit by the best Bochum/Juelich 
N$^3$LO $pp$ potential results in
a $\chi^2$/datum = 2.9 
and the worst is 22.3.

\section{Limitations of $\chi$PT}
Since $\chi$PT is a low-momentum expansion, 
we have to expect limitations concerning
its applicability.
This is demonstrated in Fig.~2, where phase
shift predictions by various NN potentials are
shown up to 1000 MeV lab.\ energy for the incident
nucleon. The figure includes one representative
of the family of the high-precision NN potentials,
namely, the CD-Bonn potential~\cite{Mac01} (solid line), which
obviously predicts the phase shifts correctly
up to the highest energies shown, even though
it was adjusted only up to 350 MeV. The same is true
for other high-precision potentials, like 
Argonne $V_{18}$~\cite{WSS95} and the Nijmegen
potentials~\cite{Sto94}.
On the other hand, the chiral NN potentials at
order N$^3$LO (dashed line Ref.~\refcite{EM03}
and dotted line Ref.~\refcite{EGM05})
do not make any reasonable predictions beyond
about 300 MeV lab.\ energy. This is, of course,
not unexpected
since $\chi$PT applies only for momenta
$Q \ll \Lambda_\chi \approx 1$ GeV,
which is enforced by the regulator Eq.~(1)
where a typical choice for the cutoff parameter
is $\Lambda \approx 500$ MeV.
Thus, chiral potentials are reliable only
for CMS momenta 
$p,p' \lesssim 2.2$ fm$^{-1}$.
A Fermi momentum 
$k_F \approx 2.2$ fm$^{-1}$
is equivalent to a nuclear matter density
$\rho \approx 4\rho_0$ where $\rho_0$
denotes nuclear matter saturation density.
Nuclear matter
calculations in which chiral two-nucleon potentials are
applied can be found in Refs.~\refcite{AS03,Li06}. 
Note, however, that for ``complete'' calculations
also the chiral 3NF must be included.
In any case, based upon the above arguments,
one may trust the $\chi$PT approach
up to densities around 4$\rho_0$.
In contrast, relativistic meson theory
can be trusted to very high momenta (cf.\
the CD-Bonn curve in Fig.~2)
and densities equivalent to those high momenta.

\begin{figure}[t]
\vspace*{-1.3cm}
\psfig{file=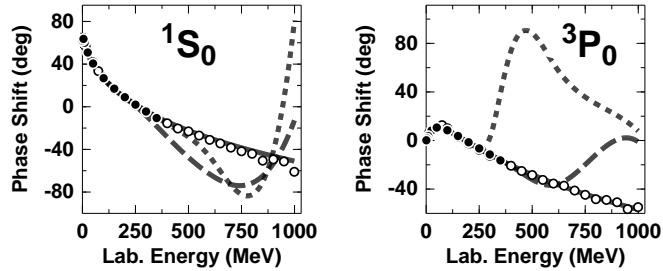,width=120mm}
\vspace*{-9.80cm}
\caption{$np$ phase shifts of the $^1S_0$ and $^3P_0$ partial
waves for lab.\ energies up to 1000 MeV.
The solid curve shows the phase shifts
predicted by the CD-Bonn potential~\cite{Mac01}.
Note that this curve is hardly visible because it
agrees with the data and, thus, is buried under the symbols
representing the data.
The dashed and the dotted lines are
the predictions by the
N$^3$LO chiral potentials constructed by the
Idaho~\cite{EM03} 
and 
the Bochum/Juelich~\cite{EGM05} groups, respectively.
Solid dots represent the Nijmegen
multienergy $np$ phase shift analysis~\cite{Sto93} and 
open circles the
GWU/VPI single-energy $np$ analysis SM99~\cite{SM99}.
}
\end{figure}

\section{Conclusions}

The theory of nuclear forces has made great strides
since the turn of the millennium.
Nucleon-nucleon potentials have been developed that are based 
on proper theory (EFT for low-energy QCD) 
and are of high-precision, at the same time. 
Moreover, the theory generates
two- and many-body forces on an equal footing
and provides a theoretical explanation for
the empirically known fact that 2NF $\gg$ 3NF $\gg$ 4NF.

At N$^3$LO~\cite{EM02,EM03}, the accuracy can be achieved that
is necessary and sufficient for microscopic nuclear structure.
First calculations applying the N$^3$LO
NN potential~\cite{EM03} in the (no-core) shell model 
\cite{Cor02,NC04,FNO05,Var05},
the coupled cluster formalism
\cite{Kow04,DH04,Wlo05,Dea05,Gou06},
and the unitary-model-operator approach~\cite{FOS04}
have produced promising results. 

The 3NF at NNLO is known~\cite{Kol94,Epe02b} and has been
applied in few-nucleon reactions~\cite{Epe02b,Erm05,Wit06}
as well as the structure of light nuclei~\cite{Nog04,Nog06,Nav07}. 
However, the famous `$A_y$ puzzle' of nucleon-deuteron
scattering is not resolved by the 3NF at NNLO. Thus, the 
most important outstanding issue is the 3NF at N$^3$LO,
which is under construction.

\section*{Acknowledgments}

This work was supported in part by the U.S.
National Science Foundation under Grant No.~PHY-0099444.

\end{document}